\documentclass[aps,pra,twocolumn,amsmath,amssymb,superscriptaddress]{revtex4}
\usepackage[latin2]{inputenc}
\usepackage{amsmath}
\usepackage{graphicx}
\usepackage{multirow}
\usepackage{dcolumn}
\usepackage{epstopdf}
\usepackage{color}

\newcommand{\ba}{\begin{eqnarray*}}
\newcommand{\ea}{\end{eqnarray*}}
\newcommand{\baa}{\begin{eqnarray}}
\newcommand{\eaa}{\end{eqnarray}}
\def\bar{\begin{array}}
\def\ear{\end{array}}
\def\LB{\left(}
\def\RB{\right)}

\def\f{\frac}

\def\nn{\nonumber}

\def\K{\mathcal{K}}
\def\M{\mathcal{M}}
\def\ve{\varepsilon}

\begin{document}

\title{Asymptotic analysis of the Berry curvature in the $E\otimes e$ Jahn-Teller model} 

\author{Ryan Requist}
\email{rrequist@mpi-halle.mpg.de}
\affiliation{
Max Planck Institute of Microstructure Physics, Weinberg 2, 06114 Halle, Germany 
}
\author{C\`esar R. Proetto}
\affiliation{
Centro At\'omico Bariloche and Instituto Balseiro, 8400 San Carlos de Bariloche, R\'io Negro, Argentina
}
\author{E. K. U. Gross}
\affiliation{
Max Planck Institute of Microstructure Physics, Weinberg 2, 06114 Halle, Germany
}
\affiliation{
Fritz Haber Center for Molecular Dynamics, Institute of Chemistry, The Hebrew University of Jerusalem, Jerusalem 91904 Israel
}

\date{\today}

\begin{abstract}
The effective Hamiltonian for the linear $E\otimes e$ Jahn-Teller model describes the coupling between two electronic states and two vibrational modes in molecules or bulk crystal impurities.  While in the Born-Oppenheimer approximation the Berry curvature has a delta function singularity at the conical intersection of the potential energy surfaces, the exact Berry curvature is a smooth peaked function.  Numerical calculations revealed that the characteristic width of the peak is $\hbar \K^{1/2}/g\M^{1/2}$, where $\M$ is the mass associated with the relevant nuclear coordinates, $\K$ is the effective internuclear spring constant and $g$ is the electronic-vibrational coupling.  This result is confirmed here by an asymptotic analysis of the $\M\rightarrow\infty$ limit, an interesting outcome of which is the emergence of a separation of length scales.  Being based on the exact electron-nuclear factorization, our analysis does not make any reference to adiabatic potential energy surfaces or nonadiabatic couplings.  It is also shown that the Ham reduction factors for the model can be derived from the exact geometric phase.

\end{abstract}

\maketitle

\section{Introduction}

Some polyatomic molecules display a peculiar type of cyclic vibrational motion in which the molecule passes through a sequence of distorted configurations that are equivalent modulo rigid rotations of the whole molecule.  Since no real rotation takes place, such motion is called \textit{pseudorotation}.  A similar phenomenon occurs for bulk crystal impurities, where the local crystal structure can distort in various symmetry-equivalent ways, e.g.~the environment of an impurity at an octahedral site can deform tetragonally in $x$, $y$ or $z$ directions.  If the potential barriers between equivalent minimum energy distorted structures are low enough, the rapid interconversion between them, known as the \textit{dynamical Jahn-Teller effect}, restores the higher symmetry of the undistorted state.

If one tracks the electronic Born-Oppenheimer (BO) wave function along a closed pseudorotational path in nuclear coordinate space, choosing its phase so that it always remains real-valued, one finds that it changes sign after one complete cycle if the path encircles a conical intersection of the adiabatic potential energy surfaces.  The electronic wave function chosen this way is therefore a double-valued function of the nuclear coordinates.  The sign change, known as the Longuet-Higgins phase \cite{longuet-higgins1958,herzberg1963}, is a special case of the Berry phase \cite{berry1984,mead1979}, and its effects
are observable in the vibrational spectroscopy of pseudorotating molecules \cite{vonbusch1998,delacretaz1986} and electron paramagnetic resonance \cite{coffman1965,coffman1966,ham1972,englman1972,ham1987} and optical \cite{sturge1967,sturge1970} spectroscopy of transition metal impurities in bulk crystals.  Evidence for the dynamical Jahn-Teller effect in the excited states of the nitrogen-vacancy center in diamond has been reported \cite{davies1979,fu2009,abtew2011,plakhotnik2015,ulbricht2016}, making the Longuet-Higgins phase relevant to its optical properties.  Recent theoretical work has explored the sign change in the bound states of small molecules by \textit{ab initio} and model calculations \cite{kendrick1997,allen2005,perebeinos2005,baer2010,lee2013,joubertdoriol2013,min2014,englman2015,requist2016a}.  

To see that the Longuet-Higgins phase is a special case of the Berry phase, one can change from the gauge in which the electronic wave function $\tilde{\Phi}_R^{\rm BO}(r)$ is real 
and double-valued to one in which it is complex and single-valued, i.e.~$\Phi_R^{\rm BO}(r) = \tilde{\Phi}_R^{\rm BO}(r) \mathrm{exp}(\f{i}{\hbar} \int A_{\mu} dR_{\mu})$, 
where the vector potential $A_{\mu}$ is chosen so that the Dirac phase factor cancels the sign change \cite{mead1979}.  The Longuet-Higgins phase for the pseudorotational path $\mathcal{C}$ is then recovered by evaluating the Berry phase formula
\begin{align}
\gamma^{\rm BO} =  \f{1}{\hbar} \oint_{\mathcal{C}} \mathrm{Im} \langle \Phi^{\rm BO}_R | \partial_{\mu} \Phi^{\rm BO}_R \rangle dR_{\mu} {,} \label{eq:gamma:LH}
\end{align}
where $\partial_{\mu}=\partial/\partial R_{\mu}$, $R=\{R_{\mu}\}$ denotes the set of nuclear coordinates and the inner product is taken with respect to electronic coordinates $r=(\mathbf{r}_1,\mathbf{r}_2,\ldots)$ only.  Throughout the paper, an implicit sum over repeated indices is assumed.  Equation (\ref{eq:gamma:LH}) can be transformed to an integral over the Berry curvature $B_{\mu\nu}^{\rm BO} = 2\hbar\, \mathrm{Im} \langle \partial_{\mu} \Phi_R^{\rm BO}  | \partial_{\nu} \Phi_R^{\rm BO} \rangle$,
\begin{align}
\gamma^{\rm BO} = \f{1}{\hbar} \iint_{\mathcal{S}} B_{\mu\nu}^{\rm BO} dR_{\mu} dR_{\nu} {,}
\label{eq:gamma:LH:2}
\end{align}
where $\mathcal{S}$ is a surface bounded by $\mathcal{C}$.  In the BO approximation, the Berry curvature is zero except at conical intersections of the adiabatic potential energy surfaces, where it has delta function singularities.

In this paper, we consider the Berry curvature calculated with the conditional electronic wave function from the exact electron-nuclear factorization \cite{hunter1975,gidopoulos2014,abedi2010} instead of the BO wave function and study its asymptotic behavior in the large mass limit of the $E\otimes e$ Jahn-Teller model.  Jahn-Teller models, which describe the coupling between electrons and vibrations, were originally introduced to explain the instability of electronically-degenerate nonlinear polyatomic molecules to static symmetry-lowering distortions \cite{jahn1937,vanvleck1939}.  Analytical results for various Jahn-Teller models have been obtained using perturbative and asymptotic approximations \cite{longuet-higgins1958,mclachlan1961,slonczewski1967,ham1968,obrien1971,voronin1976,obrien1976,karkach1978,obrien1979,judd1979,darlison1987}, a canonical transformation method in second quantization \cite{alper1970,wagner1972,sigmund1973,shultz1976,barentzen1978,reik1987,szopa1997} and approximations based on coherent states \cite{judd1975,chancey1984,dunn2001}.  

The motivation for a detailed asymptotic analysis of the Berry curvature comes from a recent nonadiabatic generalization of density functional theory \cite{requist2016b}, where the exchange-correlation energy is a functional of the Berry curvature in addition to the density.  Unlike standard density functional theory \cite{hohenberg1964,kohn1965}, which depends on the BO approximation, nonadiabatic density functional theory is an exact theory of electrons and nuclei.  Having an explicit formula for the Berry curvature in a representative model system, as well as an understanding of how it depends on parameters such as the nuclear mass and electronic-vibrational coupling, might yield insights into the Berry curvature dependence of the functional.  

However, most of the analytical studies cited above have focused on approximating the eigenvalue spectrum, as needed to explain the unique spectroscopic signatures of Jahn-Teller systems, while the Berry curvature is a property of the wave function.  Our purpose here is to revisit the problem using an exact factorization-based analysis that it is better suited to evaluating the Berry curvature. We obtain intuitive and compact formulas for the Berry curvature, nuclear wave function and nonadiabatic contributions to the potential energy surface that are accurate for large nuclear mass.  Unlike traditional analyses that take the BO approximation as a starting point, our calculations make no reference to the adiabatic potential energy surfaces and nonadiabatic couplings.  

Two key aspects of our analysis are a transformation to coupled nonlinear differential equations and the emergence of a separation of length scales.  
These two features justify our use of different approximations in different regions of nuclear configuration space.  
The separation of length scales may be of interest beyond the $E\otimes e$ Jahn-Teller model because it suggests that the Berry curvature, as a function that is nonzero only in the immediate neighborhood of the conical intersection, might have effectively higher symmetry than other variables, e.g.~the nuclear wave function.  Such emergent symmetry might be relevant to understanding the structure of functionals in nonadiabatic density functional theory.  Our analysis is nonperturbative, as the $\M\rightarrow \infty$ limit is a singular limit of the Schr\"odinger equation.

In Sec.~\ref{sec:exact}, we review the definition of the Berry curvature beyond the BO approximation.  In Sec.~\ref{sec:Jahn-Teller}, we introduce the linear $E\otimes e$ Jahn-Teller Hamiltonian and derive the coupled electronic and nuclear Schr\"odinger equations within the exact factorization scheme.  Approximations to the nuclear wave function and the Berry curvature are derived from an asymptotic analysis in Sec.~\ref{sec:asymptotic}.  Non\-adiabatic terms in the potential energy surface are investigated in Sec.~\ref{sec:surface}.  Finally, a relationship between Ham reduction factors and the beyond-BO molecular Berry phase is derived in Sec.~\ref{sec:Ham}.

\section{\label{sec:exact} Exact Berry curvature}

Since the BO Ansatz $\Phi^{\rm BO}_R(r) \chi^{\rm BO}(R)$ is an approximation to the true electron-nuclear wave function $\Psi(r,R)$, the Longuet-Higgins phase only approximately characterizes the latter \cite{gidopoulos2014} and is actually an artifact in some cases \cite{min2014}.  An exact molecular geometric phase can be defined by replacing $\Phi^{\rm BO}_R(r)$ in Eq.~(\ref{eq:gamma:LH}) by the conditional electronic wave function $\Phi_R(r) = \Psi(r,R)/\chi(R)$ derived within the exact factorization scheme, where $\chi(R) = e^{iS(R)} \left[ \int |\Psi(r,R)|^2 dr \right]^{1/2}$ is the nuclear wave function with arbitrary phase $S(R)$ \cite{gidopoulos2014,abedi2010,min2014,requist2016a}.  Calculations for a model pseudorotating triatomic molecule found that the exact geometric phase deviates from the Longuet-Higgins phase of $\pi$ due to non\-adiabatic effects near the conical intersection of the adiabatic potential energy surfaces \cite{requist2016a}.  To understand these deviations, we write the molecular geometric phase as a surface integral
over the exact Berry curvature $B_{\mu\nu} = 2\hbar\, \mathrm{Im} \langle \partial_{\mu} \Phi_R  | \partial_{\nu} \Phi_R \rangle$, i.e.
\begin{align}
\gamma = \f{1}{\hbar} \iint_{\mathcal{S}} B_{\mu\nu} dR_{\mu} dR_{\nu} {.}
\label{eq:geomphase:exact}
\end{align}
If the coordinates $R_{\mu}$ are chosen so that the conical intersection lies in the $(R_1,R_2)$ plane, the so-called ``branching plane,'' then the relevant elements of the Berry curvature are $B_{12}$ and $B_{21}=-B_{12}$. While $B_{12}$ is a delta function in the BO approximation, an exact calculation shows that the delta function gets broadened into a smooth peaked function while its integrated weight is preserved. 
Hence, for a finite surface $\mathcal{S}$, $\gamma$ will generally be less then $\gamma^{BO}$.  The peak in $B_{12}$ is centered on the conical intersection and has a characteristic width of order $\hbar \K^{1/2}/g\M^{1/2}$ for large $\M$, where $\M$ is the nuclear mass, $\K$ is the effective spring constant of the internuclear repulsion and $g$ is the electronic-vibrational coupling \cite{requist2016a}.  The exact Berry curvature must reduce to the adiabatic Berry curvature as $\M\rightarrow\infty$, but it is a nontrivial problem to determine its functional form as it sharpens and contracts to a delta function in this limit.

\section{Linear $E\otimes e$ Jahn-Teller model \label{sec:Jahn-Teller}}

\begin{figure}[t]
\centering
     \includegraphics[width=0.9\columnwidth]{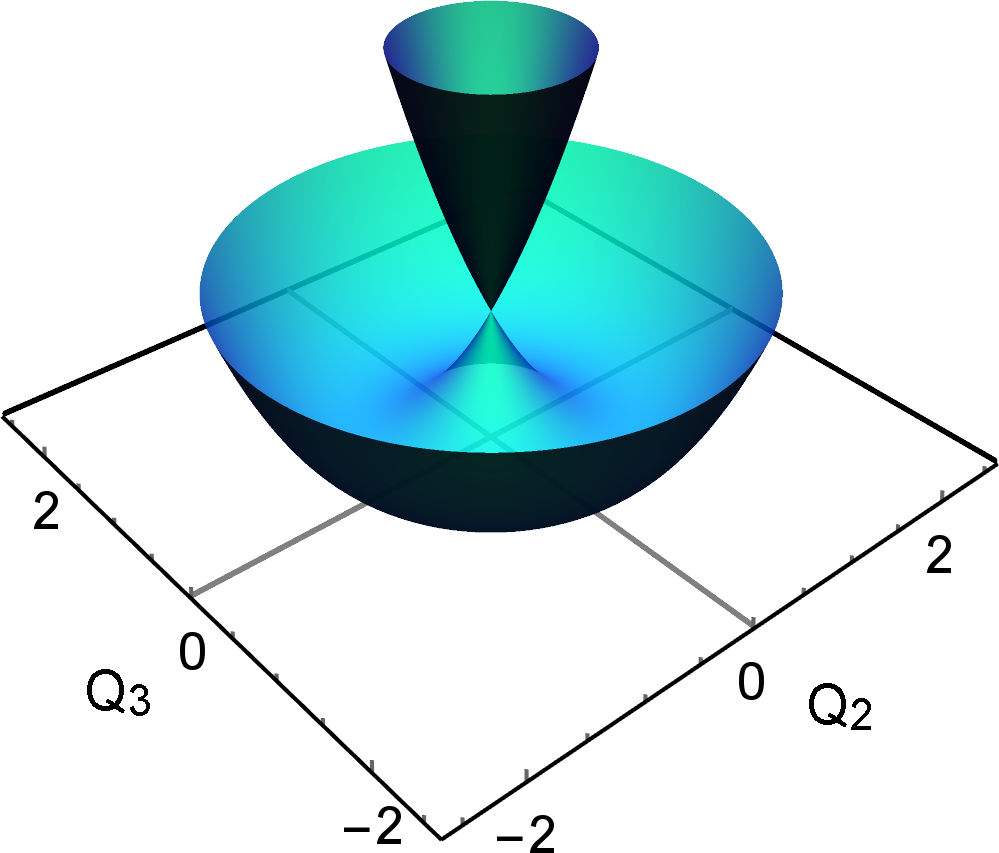} 
     \caption{Adiabatic potential energy surfaces for the linear $E\otimes e$ Jahn-Teller model with respect to vibrational normal mode coordinates $Q_2$ and $Q_3$.}
     \label{fig:Epsilon}
\end{figure}

Some molecules and bulk crystal impurities can be approximated by an effective Hamiltonian, called a Jahn-Teller or vibronic coupling model \cite{jahn1937,vanvleck1939}, comprising just a few relevant electronic states and vibrational modes. The simplest such model in which one observes a nontrivial Berry curvature, the linear $E\otimes e$ Jahn-Teller model, consists of an electronic doublet $E$ linearly coupled to a two-fold degenerate vibrational mode $e$.  Its cylindrically-symmetric adiabatic potential energy surfaces are shown in Fig.~\ref{fig:Epsilon} as a function of the vibrational normal mode coordinates, denoted $Q_2$ and $Q_3$.  
The conical intersection at the origin occurs for a high symmetry nuclear configuration, e.g.~the equilateral geometry of a triatomic molecule.  Due to the electronic-vibrational coupling, any static distortion away from the origin lifts the electronic degeneracy.  

The Hamiltonian of the $E\otimes e$ Jahn-Teller model is
\begin{align}
\hat{H} = -\f{\hbar^2}{2\M} \LB \f{d^2}{dQ_2^2} + \f{d^2}{dQ_3^2} \RB + \f{\K}{2} \LB Q_2^2 + Q_3^2 \RB + \hat{H}_{en} {,} \label{eq:H}
\end{align}
where the linear electronic-vibrational coupling is
\begin{align}
\hat{H}_{en} = g \LB \bar{cc} Q_2 & -Q_3 \\ -Q_3 & -Q_2 \ear \RB
\label{eq:Hen}
\end{align}
in a basis of electronic states $\{ |u\rangle, |g\rangle \}$ that are odd/even with respect to $Q_3$ reflection. Defining cylindrical coordinates $Q=\sqrt{Q_2^2+Q_3^2}$ and $\eta=\tan^{-1}(Q_3/Q_2)$ and applying the unitary transformation $\hat{U}= (( i , -i ), ( 1 , 1 ))/\sqrt{2}$ yields the Hamiltonian in the basis of current-carrying electronic states $|\pm\rangle = (|g\rangle\pm i |u\rangle)/\sqrt{2}$ as 
\begin{align}
\hat{H}^{\prime} = -\f{\hbar^2}{2\M} \LB \f{1}{Q} \f{d}{dQ} Q \f{d}{dQ} + \f{1}{Q^2} \f{d^2}{d\eta^2} \RB + \f{\K}{2} Q^2 + \hat{H}_{en}^{\prime} \label{eq:Hprime}
\end{align}
with 
\begin{align}
\hat{H}_{en}^{\prime} = \hat{U}^{\dag} \hat{H}_{en} \hat{U} =  g \LB \bar{cc} 0 & -Q e^{-i\eta} \\ -Q e^{i\eta} & 0 \ear \RB {.} 
\end{align}

To simplify the analysis, we will exploit the electronic-vibrational (vibronic) symmetry of the model \cite{moffitt1957a,moffitt1957b,longuet-higgins1958}.  First, define an operator $\hat{\tau}_z$ such that
\begin{align}
\hat{\tau}_z |\pm\rangle &= \pm|\pm\rangle {.}
\end{align}
The electronic angular momentum operator $\hat{l}_z=\f{\hbar}{2}\hat{\tau}_z$ has eigenvalues $l=\pm \hbar/2$.  We then define a pseudorotational angular momentum operator $\hat{L}_z=-i \hbar\partial/\partial \eta$ and the total angular momentum operator
\begin{align}
\hat{J}_z = \hat{L}_z+\hat{l}_z {.}
\end{align}
Since $\hat{J}_z$ commutes with the Hamiltonian, all states can be labeled by the quantum number $j=m+l$ which takes the values $\pm 1/2, \pm 3/2, \pm 5/2, \ldots$.  Only states with the same value of $j$ are coupled.  $\hat{J}_z$ is the generator corresponding to the rotational symmetry of the model.  The general form of a state with quantum number $j$ is
\begin{align}
|\Psi_j(Q,\eta)\rangle =\LB \bar{c} a_j(Q) e^{i(j-\f{1}{2})\eta)} \\ b_j(Q) e^{i(j+\f{1}{2})\eta)} \ear \RB {.}
\end{align}
The ground state is a $j=\pm 1/2$ doublet.  Our calculations will be made for the $j=1/2$  state $|\Psi\rangle = a|+\rangle + be^{i\eta}|-\rangle$, where here and hereafter we suppress the subscript $j$.  The Schr\"odinger equation becomes 
\begin{align}
 \left[ -\f{\hbar^2}{2\M} \LB \f{1}{Q} \f{d}{dQ} Q \f{d}{dQ} \RB + \f{\K}{2} Q^2 \right] &\LB \bar{c} a \\ b  \ear \RB \nn \\
 + \LB \bar{cc} 0 & -gQ \\ -gQ  & \hbar^2/2\M Q^2 \ear \RB &\LB \bar{c} a \\ b  \ear \RB = E \LB \bar{c} a \\ b \ear \RB  {,}\label{eq:schroedinger}
\end{align}
which is a linear system of differential equations for the functions $a=a(Q)$ and $b=b(Q)$, which are additionally required to satisfy the normalization condition
\begin{align}
\int_0^{2\pi} d\eta \int_0^{\infty} (a^2 + b^2 ) Q dQ = 1 {.} \label{eq:normalization}
\end{align}

Using the exact factorization scheme \cite{hunter1975,gidopoulos2014,abedi2010}, we define the nuclear wave function
\begin{align}
\chi=\chi(Q) &= \left[\int \Psi^*(r,R) \Psi(r,R) dr \right]^{1/2} \nn \\
&=  \sqrt{a^2+b^2} 
\end{align}
and the conditional electronic wave function
\begin{align}
|\Phi_R\rangle &= \f{|\Psi(Q,\eta)\rangle}{\chi(Q)} = \LB \bar{l} \cos\f{\theta}{2} \\[0.2cm] \sin\f{\theta}{2} e^{i\varphi} \ear \RB  {,} \label{eq:PhiR}
\end{align}
where the subscript $R$ denotes a parametric dependence on the nuclear coordinates $R=(Q,\eta)$ and $|\Phi_R\rangle$ has been expressed in terms of the Bloch sphere angles 
\begin{align}
\theta=\theta(Q) = 2 \tan^{-1} \f{b}{a} \quad \textrm{and} \quad \varphi=\eta {.}
\end{align}
The exact factorization scheme converts the original full Schr\"odinger equation into separate electronic and nuclear Schr\"odinger equations \cite{gidopoulos2014}.  For the present model, the nuclear equation is found to be 
\begin{align}
-\f{\hbar^2}{2\M} \left[ \f{1}{Q} \f{d}{dQ} Q \f{d}{dQ} - \f{1}{Q^2} \sin^4\f{\theta}{2} \right] \chi + \mathcal{E}(Q) \chi = E \chi {,} \label{eq:chi}
\end{align}
where the second term in the brackets is $A_{\eta}^2(Q)$ with $A_{\eta}(Q)=\hbar\, \mathrm{Im} \langle \Phi_R | \partial_{\eta} \Phi_R \rangle$.  The scalar potential $\mathcal{E}(Q)$ is 
\begin{align}
\mathcal{E}(Q) &= \f{\K}{2} Q^2 - g Q\sin\theta + \mathcal{E}_{\rm geo}(Q) {,} \label{eq:Epsilon}
\end{align}
where 
\begin{align}
\mathcal{E}_{\rm geo}(Q) &= \f{\hbar^2}{2\M} \left[ \f{1}{4} \LB \f{d\theta}{dQ} \RB^2 + \f{\sin^2\theta}{4Q^2} \right]  \label{eq:EpsilonGeo}
\end{align}
is a term of geometric origin \cite{requist2016a}.  The $\theta$-dependence of $\mathcal{E}$ accounts for nonadiabatic effects, as will be discussed below.  Since $|\Phi_R\rangle$ is fully determined by $\theta$ and $\varphi$, and $\varphi$ is a known function of $R$, the electronic Schr\"odinger equation can be replaced by the following differential equation: 
\begin{align}
Q^2 \f{d^2\theta}{dQ^2} + \LB 1+ Q \f{d\log|\chi|^2}{dQ} \RB Q \f{d\theta}{dQ} -\sin\theta& \nn \\+ \f{4g\M}{\hbar^2} Q^3 \cos\theta& = 0 {,}
\label{eq:theta}
\end{align}
which can be derived from the Euler-Lagrange equation for the stationarity of $E$ or directly from the Schr\"odinger equation.  We observe that the $d^2\theta/dQ^2$ and $d\theta/dQ$ terms come from the first term of $\mathcal{E}_{\rm geo}$, the $\sin\theta$ term comes from the sum of $A_{\eta}^2$ and the second term of $\mathcal{E}_{\rm geo}$, and the last term comes from the coupling $gQ \sin\theta$.

We have thus transformed the original Schr\"odinger equation, Eq.~(\ref{eq:schroedinger}), into a pair of coupled nonlinear equations, one for the nuclear variable $\chi$ and one for the electronic variable $\theta$.  This transformation can be realized as the simple change of variables $(a,b)\rightarrow (\chi,\theta)$.  

To obtain the ground state, we need to solve Eqs.~(\ref{eq:chi}) and (\ref{eq:theta}) subject to the inner and outer boundary conditions, $\theta(0)=0$ and $\theta(\infty)=\pi/2$, and the normalization condition in Eq.~(\ref{eq:normalization}).  The inner boundary condition is necessary in order for the $j=1/2$ state to have bounded energy, since the rotational energy 
\begin{align}
\Big< \Psi \Big| -\f{\hbar^2}{2\M Q^2} \f{d^2}{d\eta^2} \Big| \Psi \Big> = \int_0^{\infty} dQ \f{\pi\hbar^2\chi^2}{\M Q}  \sin^2\f{\theta}{2}
\end{align}
diverges unless either $\chi(0)=0$ or $\theta(0)=0$.  It will later be shown that $\chi(0)\neq 0$, so we must have $\theta(0)=0$.  The outer boundary condition is necessary in order to obtain the ground state: since Eq.~(\ref{eq:schroedinger}) reduces to the BO equation in the $Q\rightarrow\infty$ limit, the solution must converge to the lower energy BO state, implying $a(\infty)=b(\infty)$ and hence the boundary condition $\theta(\infty)=\pi/2$. 

\begin{figure}[t]
\centering
     \includegraphics[width=\columnwidth]{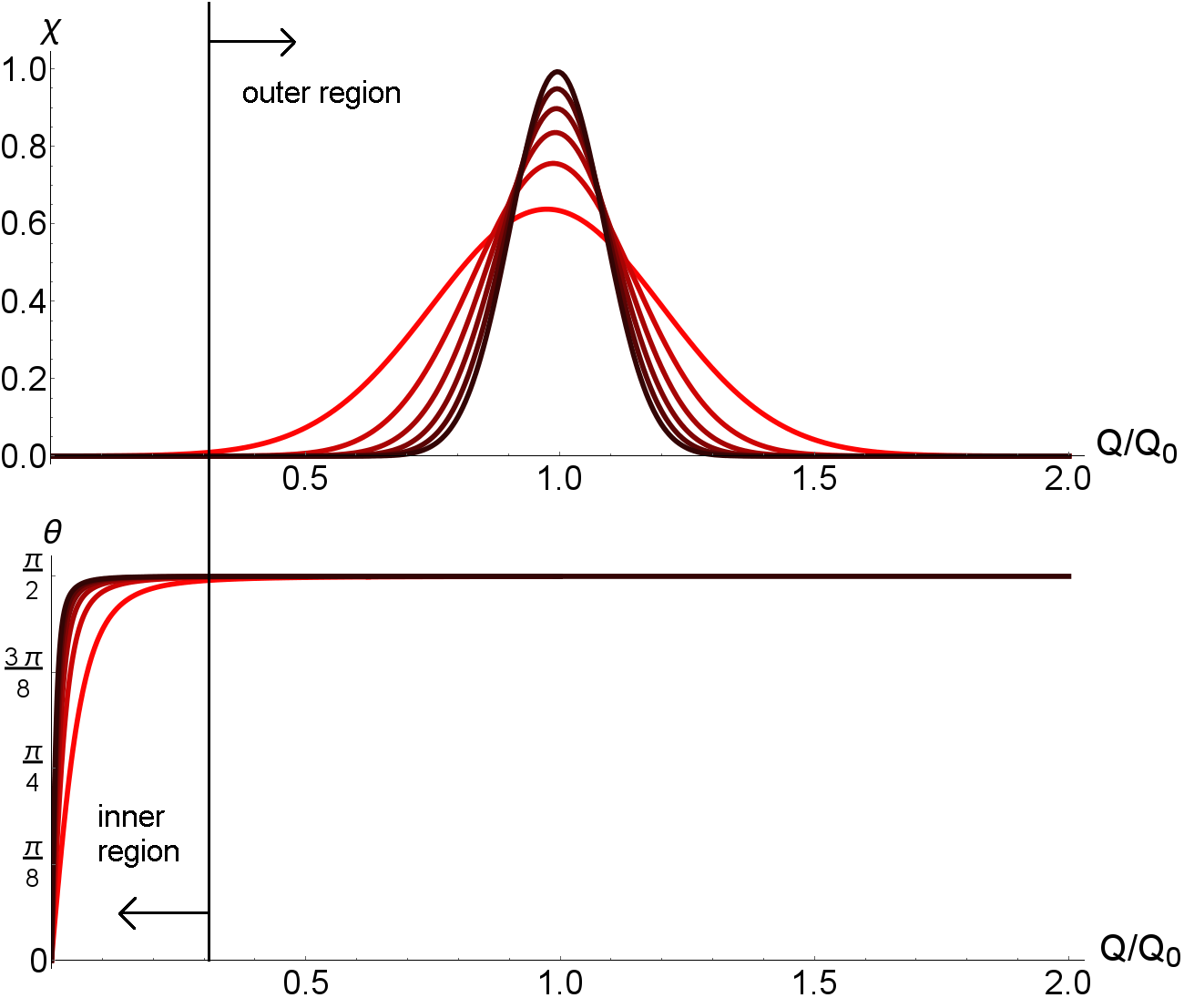} 
     \caption{Nuclear wave function $\chi(Q)$ and the electronic variable $\theta(Q)$ for increasing values of $\mathcal{M}$ [light red to dark red].  The minimum of the adiabatic potential energy surface occurs at $Q/Q_0=1$.}
     \label{fig:chitheta-vs-Q}
\end{figure}

Even before solving the differential equations, we can evaluate the Berry curvature and molecular Berry phase in terms of $\theta(Q)$ and discuss the consequences of the inner and outer boundary conditions.  Since the nuclear configuration space is two-dimensional, the Berry curvature can be represented as a $2\times 2$ matrix.  Since it is an antisymmetric matrix, it is completely determined by the single element
\begin{align}
B_{Q_2Q_3} = \f{1}{Q} B_{Q\eta} = \f{\hbar}{2} \f{1}{Q} \sin\theta \f{d\theta}{dQ} {.} \label{eq:Berrycurvature}
\end{align}
The electronic variable $\theta(Q)$ determines the shape of the Berry curvature as a function of $Q$.  Figure~\ref{fig:chitheta-vs-Q} shows that $\theta(Q)$ develops a sharp step at $Q=0$ in the $\M\rightarrow\infty$ limit.  As $\theta(Q)$ approaches a step function, the Berry curvature $ B_{Q_2Q_3}(Q)$ approaches a delta function, thus recovering the BO result.  The nonadiabatic effects captured by $\theta(Q)$ are responsible for smearing out the delta function to the smooth function $B_{Q_2Q_3}(Q)$ \cite{requist2016a}.

The molecular geometric phase for a circular path $\mathcal{C}$ with radius $Q$ in the $(Q_2,Q_3)$ plane can be evaluated according to Eq.~(\ref{eq:gamma:LH:2}) as
\begin{align}
\gamma(Q) &= \f{1}{\hbar} \int_0^{2\pi} d\eta \int_0^Q  dq B_{Q\eta}(q)   \nn \\
&= \pi \left[ 1 - \cos\theta(Q) \right] {.} \label{eq:gamma:JT}
\end{align}
The inner boundary condition $\theta(0)=0$ forces the exact geometric phase to vanish as $Q\rightarrow 0$, in contrast to the adiabatic case where the Longuet-Higgins phase remains equal to $\pi$ for any finite $Q>0$, no matter how small.  The vanishing of the geometric phase coincides with a transfer of angular momentum from nuclei to electrons as $Q\rightarrow 0$.  Since we have chosen a gauge in which $\chi$ is real, we have the identity $\langle \Phi_R | \hat{L}_z | \Phi_R \rangle + \langle \Phi_R | \hat{l}_z | \Phi_R \rangle = \hbar/2$.  The second term, $\langle \Phi_R | \hat{l}_z | \Phi_R \rangle = (\hbar/2)\cos\theta$, is the angular momentum carried by the electrons, conditional on $Q$.  Since $\theta(0)=0$, the electrons carry the full angular momentum of the state when $Q=0$.  The conditional angular momentum carried by the nuclei is directly related to the geometric phase via $\langle \Phi_R | \hat{L}_z | \Phi_R \rangle = \hbar(\gamma/2\pi)$.

As a consequence of the outer boundary condition $\theta(\infty)=\pi/2$, the exact geometric phase $\gamma$ approaches $\pi$ as $Q\rightarrow \infty$, recovering the Longuet-Higgins phase.  This proves that although the Berry curvature is spread out by nonadiabatic effects, its integral over all space, $h/2$, is conserved.  

\section{Large mass limit \label{sec:asymptotic}}

\subsection{Overview of approximation strategy}

Before beginning the calculations, we briefly summarize our strategy and introduce the length scales needed to analyze the $\M\rightarrow \infty$ limit.

The exact factorization scheme transforms the original Schr\"odinger equation into coupled nonlinear differential equations, Eqs.~(\ref{eq:chi}) and (\ref{eq:theta}), for the functions $\chi(Q)$ and $\theta(Q)$.  Since most asymptotic methods are designed for linear differential equations, it appears that the exact factorization equations will be even more difficult to approximate than the original Schr\"odinger equation.  However, a key feature of these exact factorization equations is the emergence, as $\M\rightarrow\infty$, of a separation of length scales that is not manifest in the linear equations.  As visible in Fig.~\ref{fig:chitheta-vs-Q}, $\chi(Q)$ becomes localized in the classically-allowed region near $Q=Q_0$, where $Q_0\equiv g/\K$ is the radius at which the adiabatic potential energy surface reaches its minimum, while $\theta(Q)$ is essentially constant throughout that region and only undergoes significant changes near the origin, i.e.~on a much shorter length scale.  We will denote the region near the origin where $\theta(Q)$ rises from 0 to $\pi/2$ as the inner region and all larger $Q$ as the outer region; these regions are depicted in Fig.~\ref{fig:chitheta-vs-Q}. 

In the outer region, $\theta(Q)$ is readily approximated by a slowly-varying function.  Substituting an approximate $\theta(Q)$ into the equation for $\chi(Q)$, Eq.~(\ref{eq:chi}), yields a linear differential equation that can be approximated by standard semiclassical methods.  In the inner region, an adequate zeroth-order approximation for $\chi(Q)$ and $\theta(Q)$ can be obtained by neglecting the $gQ$ and $(\K/2)Q^2$ terms in Eq.~(\ref{eq:chi}).  Matching the inner and outer approximations gives a global approximation to $\chi(Q)$, which can then be used in the equation for $\theta(Q)$.  To make these arguments more precise, we now define the relevant length scales and a dimensionless small parameter $\ve$.

The two relevant length scales in the outer region are $Q_0$ and the amplitude of zero-point motion
\begin{align}
Q_{\rm zp} = \LB \f{\hbar^2}{\K\M} \RB^{1/4} {.}
\end{align}
A dimensionless small parameter that quantifies the degree of localization of $\chi(Q)$ is the ratio 
\begin{align}
\ve = \LB \f{Q_{\rm zp}}{Q_0} \RB^2  =  \f{\hbar \K^{3/2}}{g^2 \M^{1/2}} {.}
\end{align}
The $\M\rightarrow\infty$ limit can be realized by taking the limit $\ve\rightarrow 0$.  This dimensionless parameter can be equivalently expressed as  $\ve = \hbar\Omega/2\Delta$, i.e.~the ratio of the zero-point energy $\hbar\Omega/2$ to the Jahn-Teller stabilization energy $\Delta\equiv g^2/2\K$; the fundamental frequency is $\Omega \equiv \sqrt{\K/\M}$.  Longuet-Higgins \textit{et al.}~defined a parameter $k$ quantifying the strength of electronic-vibrational coupling in the adiabatic potentials $(1/2)r^2\pm kr$, where $r$ is a dimensionless radial coordinate \cite{longuet-higgins1958}.  Since $k=\ve^{-1/2}$, their strong coupling limit $k\rightarrow\infty$ is equivalent to our $\ve\rightarrow 0$ limit.

The relevant length scale in the inner region is the characteristic length, denoted as $Q_{\rm width}$, over which $\theta(Q)$ rises from 0 to $\pi/2$.  This gives the characteristic width of the peak in the Berry curvature.   The analysis in the following section will demonstrate that 
\begin{align}
Q_{\rm width} = \f{\hbar \K^{\f{1}{2}}}{g \M^{\f{1}{2}}} {,} \label{eq:Qsmear}
\end{align}
which is consistent with the numerical results of Ref.~\cite{requist2016a}.  Since $Q_{\rm width} =\ve Q_0$ and $Q_{\rm zp} = \ve^{1/2} Q_0$, we have the hierarchy of length scales $Q_{\rm width} \ll Q_{\rm zp} \ll Q_0$.

\subsection{Asymptotic analysis in the outer region} 

To analyze the outer region, we first perform a change of variables to bring the nuclear equation to the standard form of the Wentzel-Kramers-Brillouin (WKB) method so that it can be approximated by the method of comparison equations \cite{miller1953,dingle1956,langer1937}.  After changing the independent variable to $q=Q/Q_0$, Eqs.~(\ref{eq:chi}) and (\ref{eq:theta}) become 
\begin{align}
-\ve^2 \LB \f{1}{q} \f{d}{dq} q \f{d}{dq} - \f{1}{q^2} \sin^4\f{\theta}{2} \RB \chi + \f{\mathcal{E}}{\Delta} \chi = \f{E}{\Delta} \chi \label{eq:chi:outer} 
\end{align}
and
\begin{align}
q^2 \f{d^2\theta}{dq^2} + \LB 1 + q \f{d\log|\chi|^2}{dq} \RB q\f{d\theta}{dq} - \sin\theta& \nn \\ + \f{4}{\ve^2} q^3 \cos\theta &= 0 \label{eq:theta:outer}
\end{align} 
with 
\begin{align}
\f{\mathcal{E}}{\Delta} = q^2 - 2 q \sin\theta + \ve^2 \left[ \f{1}{4} \LB \f{d\theta}{dq} \RB^2 + \f{\sin^2\theta}{4q^2} \right] {.}
\end{align}
Next changing the dependent variable to $\mu = q^{1/2} \chi$, the nuclear equation becomes 
\begin{align}
\f{d^2\mu}{dq^2} + \f{1}{\ve^2} \left[ \f{E}{\Delta} - \f{\mathcal{E}}{\Delta} - \f{\ve^2}{q^2} \LB \sin^4\f{\theta}{2} - \f{1}{4} \RB \right] \mu = 0 \label{eq:mu:outer} {,}
\end{align}
which is in standard WKB form.  The method of comparison equations provides an approximation that is asymptotic to the exact solution in the $\ve\rightarrow 0$ limit, but unlike the WKB solution, it is uniformly valid across both turning points, so there is no need to use connection formulas to relate the solutions in classically allowed and classically forbidden domains.  Having a uniform approximation is an advantage if one needs to evaluate integrals over the solutions, as we do in Sec.~\ref{sec:Ham}.
 
Although Eq.~(\ref{eq:mu:outer}) is linear in $\mu$, it depends nonlinearly on $\theta$ through $\mathcal{E}$ and the $\sin^4(\theta/2)$ term.  To see how to approximate $\theta$ in the outer region, consider Eq.~(\ref{eq:theta:outer}) and recall the outer boundary condition $\theta(\infty)=\pi/2$.  Since $\theta(q)$ is approximately constant, a dominant balance \cite{bender1999} is achieved by neglecting the first two terms in Eq.~(\ref{eq:theta:outer}).  Hence, the lowest-order outer approximation is
\begin{align}
\theta_{\rm out,0} = \tan^{-1} \f{4q^3}{\ve^2} + \mathcal{O}(\ve^3) {.} \label{eq:theta0:outer}
\end{align} 
An effective potential $\mathcal{E}_{\rm eff,out}(q)$ in Eq.~(\ref{eq:mu:outer}) can be identified by combining the centrifugal potential with $\mathcal{E}(q)$.  Substituting $\theta_{\rm out,0}$ 
into $\mathcal{E}_{\rm eff,out}$ 
and expanding in $\ve$ gives
\begin{align}
\f{\mathcal{E}_{\rm eff,out}(q)}{\Delta} &= q^2 - 2q\sin\theta + \ve^2 \Bigg[ \f{1}{4} \bigg(\f{d\theta}{dq}\bigg)^2 +\f{\sin^2\theta}{4q^2} \Bigg] \nn \\
&\quad+ \ve^2 \f{\sin^4\f{\theta}{2} - \f{1}{4}}{q^2} \nn \\
&= q^2 - 2q + \f{\ve^2}{4q^2} - \f{\ve^4}{16q^5} + \mathcal{O}(\ve^6) {.}
\end{align}
Keeping only the terms up to $\mathcal{O}(\ve^2)$ corresponds to setting $\theta=\pi/2$ and gives the equation
\begin{align}
\f{d^2\mu_0}{dq^2} + \f{1}{\ve^2} p^2(q) \mu_0 = 0 \label{eq:mu0:outer} {,}
\end{align}
with 
\begin{align}
p^2(q) = \f{E}{\Delta} - q^2 + 2q - \f{\ve^2}{4q^2} {.} \label{eq:p}
\end{align}
The last term comes from the second term of $\mathcal{E}_{\rm geo}$ and the centrifugal potential $\ve^2 [\sin^4(\theta/2) - 1/4]/q^2$.

The idea behind the method of comparison equations is to choose an exactly solvable reference equation (the so-called \textit{comparison equation}) that resembles the original equation in the sense that it has the same number and type of turning points.  In the present case, we choose 
\begin{align}
\f{d^2 U}{dX^2} + \f{1}{\ve^2} P^2(X) U = 0 {;} \quad P^2(X) = 2J - X^2  {.}
\label{eq:U}
\end{align}
This is similar to Eq.~(\ref{eq:mu0:outer}) because $P^2(X)$, like $p^2(q)$, has two simple turning points.  It describes a harmonic oscillator with energy $J=\f{1}{2}(P^2+X^2)$.  The ground state is
\begin{align}
U(X) &= \f{1}{\sqrt{2\pi}Q_0} \LB \f{1}{\pi\ve}\RB^{1/4} e^{-X^2/2\ve} {,}
\label{eq:U:solution}
\end{align}
which implies the following approximation for $\mu_0(q)$:
\begin{align}
\mu_{0}(q) &= \f{N_{\rm out}}{\sqrt{2\pi}Q_0} \LB \f{1}{\pi\ve}\RB^{1/4} \LB \f{dX}{dq} \RB^{-1/2} e^{-X^2(q)/2\ve} {,}
\label{eq:mu:solution:q}
\end{align}
where $N_{\rm out}$ is a normalization constant and $X=X(q)$ is defined implicitly via \cite{miller1953}
\begin{align}
\int_{-\sqrt{2J}}^X P(X^{\prime}) dX^{\prime} &= \int_{q_1}^q p(q^{\prime})dq^{\prime} {.}\label{eq:X}
\end{align}
The lower limits of the integrals are the turning points defined by $P(-\sqrt{2J})=0$ and $p(q_1)=0$.  The right-hand side depends on the energy eigenvalue $E$, a first estimate for which can be obtained from the semiclassical Bohr-Sommerfeld quantization condition (for $n=0$) 
\begin{align}
\f{1}{\ve} \int_{q_1}^{q_2} p(q)dq = \f{\pi}{2} {.} \label{eq:Bohr-Sommerfeld}
\end{align}
With $p(q)$ given by Eq.~(\ref{eq:p}) this integral can be evaluated analytically. 
If we neglect the $\ve^2/4q^2$ term of $p(q)$, it gives 
\begin{align}
E_0 = -\Delta + \f{\hbar \Omega}{2} {,} 
\end{align}
which is simply the sum of the Jahn-Teller stabilization energy and the zero-point energy of radial motion.  Since
\begin{align}
\int_{-\sqrt{2J_{0}}}^{+\sqrt{2J_{0}}} P(X) dX = \pi J_{0} {,}
\end{align}
Eqs.~(\ref{eq:X}) and (\ref{eq:Bohr-Sommerfeld}) imply $J_0=\ve/2$.  To determine $E$ systematically to higher order, two solutions should be matched together in the classically allowed region -- one originating from a solution that decays to the left and the other from a solution that decays to the right.  However, the error in the semiclassical energy is here only $\mathcal{O}(\ve^4)$, which is small enough for our purposes. 
\begin{figure}[t]
\centering
     \includegraphics[width=\columnwidth]{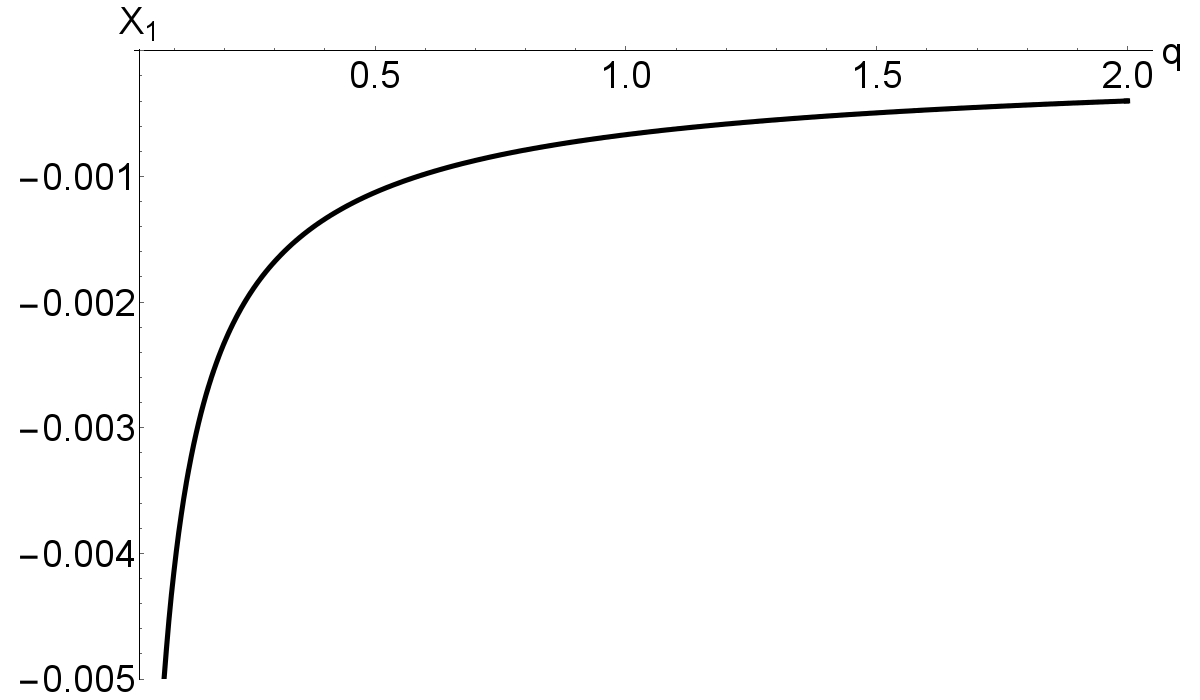} 
     \caption{The function $X_1(q)=X(q)-X_0(q)$ for $\ve^2 = 0.0025$.}
     \label{fig:X1}
\end{figure}

We now derive the function $X=X(q)$ according to the definition in Eq.~(\ref{eq:X}). The function $X(q)$ acts as a kind of deformation function, accounting for the perturbation of the harmonic oscillator wave function due to the repulsive potential $\ve^2/4q^2$.  The left-hand side of Eq.~(\ref{eq:X}) can be evaluated exactly to give
\begin{align}
\int_{-\sqrt{2J}}^X P(X^{\prime}) dX^{\prime} = \f{1}{2} X P + J \left[ \f{\pi}{2} + \tan^{-1} \f{X}{P} \right] {.} \label{eq:lefthandside}
\end{align}
The right-hand side can be evaluated in terms of elliptic functions, but the result is too lengthy to record here.  Thus, we can construct $X=X(q)$ by equating the analytical results for the left- and right-hand sides of Eq.~(\ref{eq:X}) and finding the solution numerically.  The zeroth-order approximation is $X_0(q)=q-1$.  Since the repulsive potential is a small perturbation for $q\gg \ve$, $dX/dq$ is slowly varying for large $q$.  In Fig.~\ref{fig:X1}, we plot $X(q)-X_0(q)$, which shows the small but crucial $\mathcal{O}(\ve^2)$ contribution to $X(q)$.

\begin{figure}[t!]
\centering
     \includegraphics[width=\columnwidth]{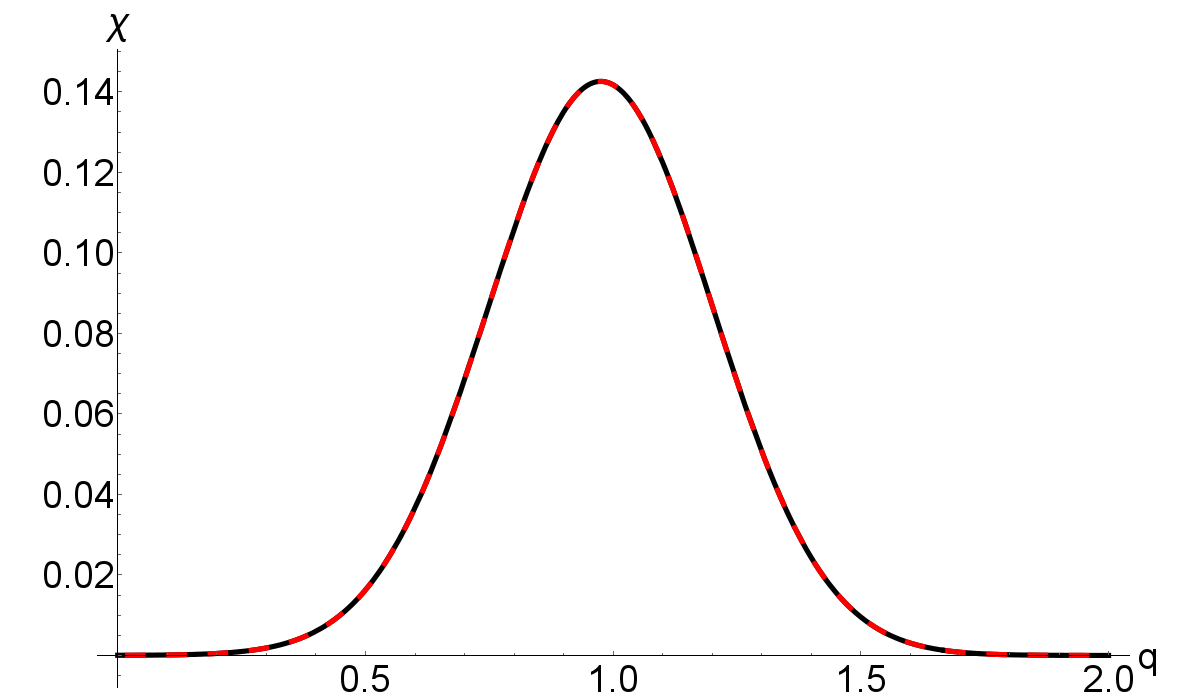} 
     \caption{The exact function $\chi_{\rm exact}(q)$ [black] and the approximation $\chi_{\rm out, 0}(q)$ [red dashed].}
     \label{fig:chi}
\end{figure}

Substituting the function $X=X(q)$ into Eq.~(\ref{eq:mu:solution:q}) and changing back to the dependent variable $\chi$ gives  
\begin{align}
\chi_{\rm out, 0}(q)=  \f{N_{\rm out}}{\sqrt{2\pi q} Q_0} \LB \f{1}{\pi\ve}\RB^{1/4} \LB \f{dX}{dq} \RB^{-1/2} e^{-X^2(q)/2\ve} {.} \label{eq:chi:out0}
\end{align}
In Fig.~\ref{fig:chi}, $\chi_{\rm out, 0}(q)$ is compared with the exact function $\chi_{\rm exact}(q)$.  
The error, $\mathcal{O}(\ve^4)$, which is too small to be seen in Fig.~\ref{fig:chi}, will be shown in Fig.~\ref{fig:chi:uniform:error}.  The approximations for $\theta_{\rm out}$ and $\chi_{\rm out}$ could be systematically improved by keeping higher powers of $\epsilon$ in Eqs.~(\ref{eq:theta:outer}) and (\ref{eq:mu:outer}).

\begin{figure}[t!]
\centering
     \includegraphics[width=\columnwidth]{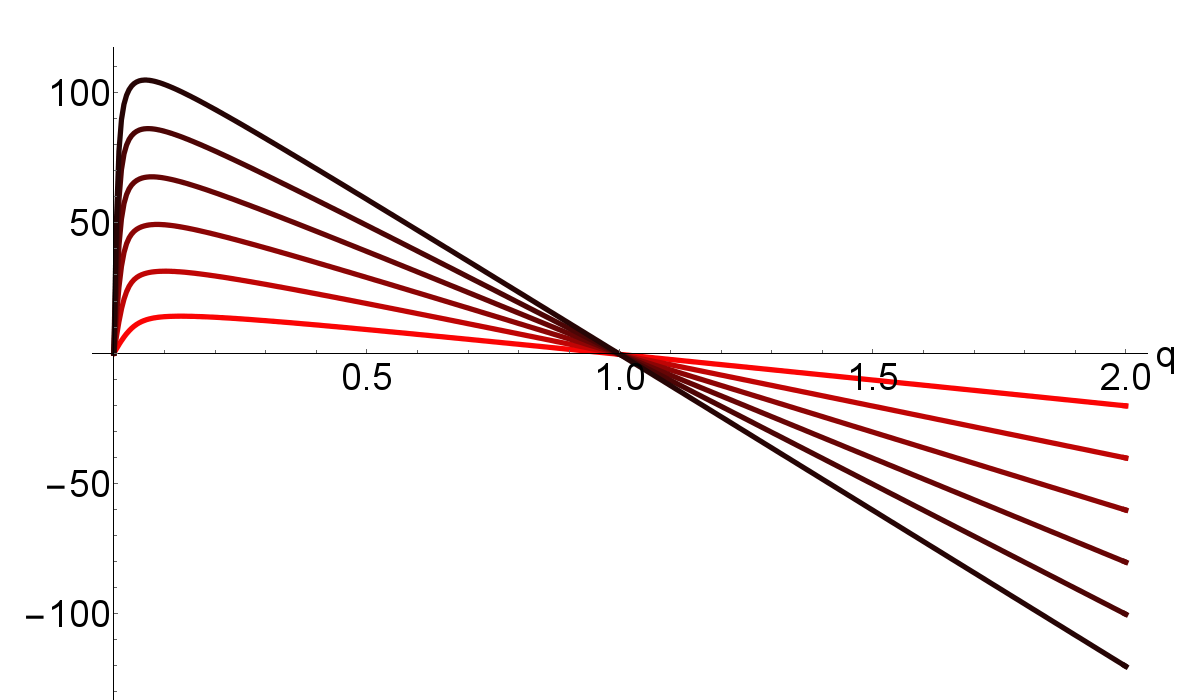} 
     \caption{The function $d\log\chi^2/dq$ is plotted for the series of values $\ve=(\f{1}{120},\f{1}{100},\f{1}{80},\f{1}{60},\f{1}{40},\f{1}{20})$ [dark red to light red].}
     \label{fig:dlogchisq}
\end{figure}

The differential equation for $\theta(q)$ depends only on $d\log\chi^2/dq$, i.e.~the relative rate of change of $\chi$ as opposed to its actual value.  Figure~\ref{fig:dlogchisq} shows the exact $d\log\chi^2/dq$ for several values of $\ve$.  The asymptotic behavior for large $q$ is $2(1-q)/\ve$, consistent with $\chi(q)$ approaching a gaussian \cite{longuet-higgins1958}
\begin{align}
\chi(q) = \f{1}{\sqrt{2\pi q}Q_0} (\pi \ve)^{-1/4} e^{-(q-1)^2/2\ve}
\end{align}
as $\ve\rightarrow 0$.  It is worth noting the following simple approximation to $d\log\chi^2/ds$:
\begin{align*}
\f{d\log\chi^2}{ds} &= \f{\beta s}{(1+s)(1+s^2)} + \bigg[ 2(1-\ve s) - \f{1}{1+s} \bigg] \f{s^2}{1+s^2} {,}
\end{align*}
which was constructed to have the correct asymptotic behavior in the limits $s\rightarrow 0$ and $s\rightarrow\infty$; $\beta$ is a constant determined in the next section.  The maximum error $0.05$ is approximately independent of $\ve$.  

\subsection{Asymptotic analysis in the inner region} 

To set up the equations in the inner region, we make a sequence of changes to the independent and dependent variables.  First, the independent variable is changed to $s=Q/Q_{\rm width}$, where $Q_{\rm width}$ is the natural length scale for the inner region.  Equations~(\ref{eq:chi}) and (\ref{eq:theta}) become 
\begin{align}
\f{1}{s} \f{d}{ds} s \f{d\chi}{ds} + \LB \f{E}{\Delta} - \f{\mathcal{E}_{\rm eff,in}}{\Delta} \RB \chi = 0   \label{eq:chi:inner}
\end{align}
with
\begin{align}
\f{\mathcal{E}_{\rm eff,in}}{\Delta} =\f{1}{4} \LB \f{d\theta}{ds} \RB^2 + \f{\sin^2\f{\theta}{2}}{s^2} - 2 \ve s \sin\theta + \ve^2 s^2 \label{eq:EpsilonEff:in}
\end{align}
and
\begin{align}
s^2 \f{d^2\theta}{ds^2} + \LB 1+ s \f{d}{ds}\log|\chi|^2 \RB s\f{d\theta}{ds}&- \sin\theta \nn \\
&+ 4 \ve s^3 \cos\theta = 0 {.}
\label{eq:theta:inner}
\end{align}
Following Ref.~\onlinecite{berry1973}, we apply the Langer transformation $s=e^x$ \cite{langer1937}, which takes the independent variable $s$ restricted to the half line to a variable $x$ whose domain is the real axis.  Defining $\tilde{\chi}(x) = \chi(e^x)$ and $\tilde{\theta}(x) = \theta(e^x)$, Eqs.~(\ref{eq:chi:inner}) and (\ref{eq:theta:inner}) transform to
\begin{align}
\f{d^2\tilde{\chi}}{dx^2} + \tilde{k}^2(x) \tilde{\chi} = 0   \label{eq:chi:inner:tilde}
\end{align}
with
\begin{align}
\tilde{k}^2(x) = \f{E}{\Delta} e^{2x} - \f{1}{4} \bigg( \f{d\tilde{\theta}}{dx} \bigg)^2 &- \sin^2\f{\tilde{\theta}}{2} \nn \\
&+ 2 \ve e^{3x} \sin\tilde{\theta} - \ve^2 e^{4x} 
\end{align}
and 
\begin{align}
\f{d^2\tilde{\theta}}{dx^2} + \f{d\log|\tilde{\chi}|^2}{dx} \f{d\tilde{\theta}}{dx} - \sin\tilde{\theta} + 4 \ve e^{3x} \cos\tilde{\theta} = 0 {.} \label{eq:theta:Langer}
\end{align}
The $\tilde{\theta}$-dependence in Eq.~(\ref{eq:chi:inner:tilde}) accounts for nonadiabatic effects.  The rate of decay of $\tilde{\chi}$ as $x\rightarrow -\infty$ is not only controlled by $(E/\Delta)e^{2x}$ but also by $(d\tilde{\theta}/dx)^2$ and $\sin^2(\tilde{\theta}/2)$, since the latter two terms will be seen to be proportional to $e^{2x}$.  Hence, nonadiabatic effects crucially influence the rate of decay of $\tilde{\chi}(x)$ as $x\rightarrow -\infty$. 

\begin{figure}[t!]
\centering
     \includegraphics[width=\columnwidth]{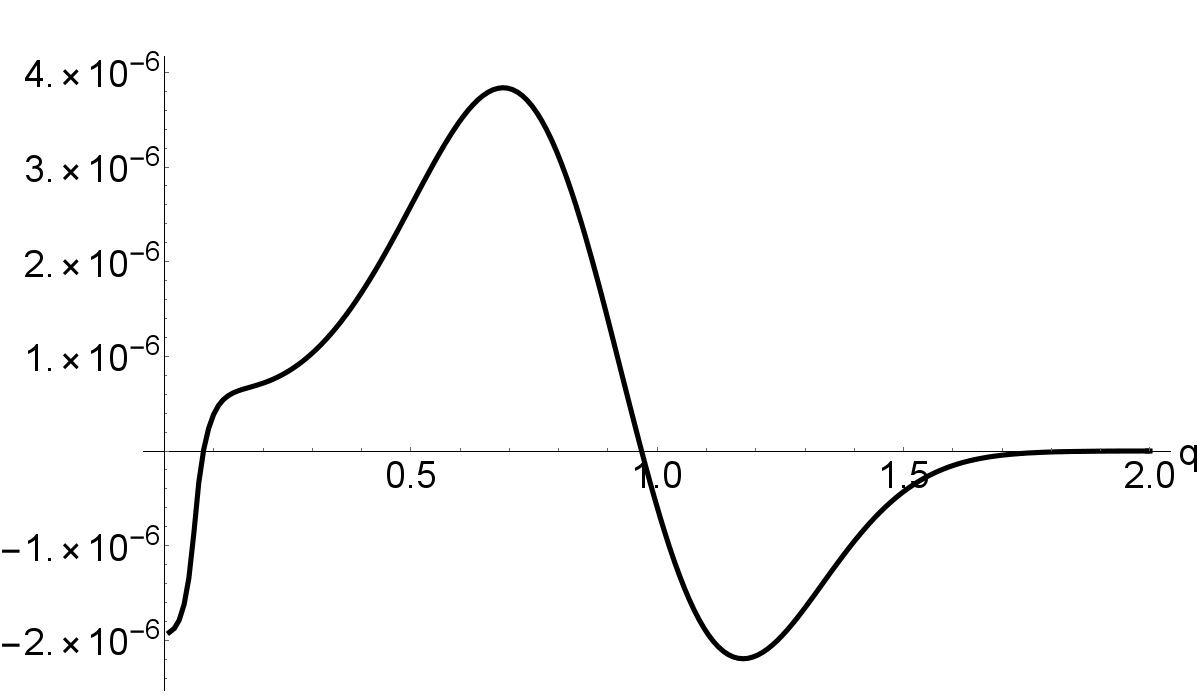} 
     \caption{The error $\chi_{\rm uniform}(q) - \chi_{\rm exact}(q)$.}
     \label{fig:chi:uniform:error}
\end{figure}

We have applied the method of comparison equations in the inner region following Ref.~\onlinecite{berry1973}; however, to obtain a simple zeroth-order approximation to Eqs.~(\ref{eq:chi:inner:tilde}) and (\ref{eq:theta:Langer}), it is more convenient to go back to the linear equations for the dependent variables $a$ and $b$.  After changing the independent variable to $s=Q/Q_{\rm width}$, Eq.~(\ref{eq:schroedinger}) transforms to
\begin{align}
\f{1}{s} \f{d}{ds} s \f{d}{ds} \LB \bar{c} a \\ b \ear \RB - \LB \bar{cc} \ve^2 s^2 & -2\ve s \\ -2\ve s & s^{-2}+\ve^2 s^2 \ear \RB \!\LB \bar{c} a \\ b \ear \RB = \f{E}{\Delta} \LB \bar{c} a \\ b \ear \RB {.} \nn
\end{align}
To zeroth-order in $\ve$, the equations for $a$ and $b$ decouple into separate equations for a free particle in cylindrical symmetry \cite{obrien1979}.  The solutions are the Bessel functions
\begin{align}
a_0(s) &= A I_0\Big(\sqrt{-\f{E}{\Delta}} s\Big) \\
b_0(s) &= B I_1\Big(\sqrt{-\f{E}{\Delta}} s\Big) {.}
\end{align}
From these solutions we can define $\chi_{\rm in,0}=\sqrt{a_0^2+b_0^2}$.  To fix the undetermined coefficients $A$ and $B$, we match the inner and outer $\chi$ and their derivatives $d\chi/ds$ at $s=1$.  Patching together the outer approximation in Eq.~(\ref{eq:chi:out0}) and the inner approximation $\chi_{\rm in,0}(s)$ defines a compact, uniform approximation $\chi_{\rm uniform}(s)$, whose $\mathcal{O}(\ve^4)$ error is shown in Fig.~\ref{fig:chi:uniform:error} for $\ve^2=0.0025$.

As anticipated from Fig.~\ref{fig:chitheta-vs-Q}, $\theta(s)$ is approximately linear for small $s$.  The proportionality constant $\alpha$ is related to the energy eigenvalue $E/\Delta$ and the rate of growth of $\chi(s)$.  We will demonstrate this by first assuming $\theta(s)=\alpha s$ in Eq.~(\ref{eq:chi:inner}) and subsequently verifying the self-consistency of this assumption.  Since the sum of the geometric and centrifugal terms in $\mathcal{E}_{\rm eff,in}(s)$ then simplifies to
\begin{align}
\f{1}{4} \bigg( \f{d\theta}{ds} \bigg)^2 + \f{\sin^2\f{\theta}{2}}{s^2} = \f{\alpha^2}{2} {,}
\end{align}
the solution to Eq.~(\ref{eq:chi:inner}) within this approximation is
\begin{align}
\chi(s) \sim I_0(\sqrt{\beta} s) {.}
\end{align}
The constant $\beta=\alpha^2/2-E/\Delta$ relates the rate of growth of $\chi(s)$ to $\alpha$ and $E/\Delta$.  Neglecting the $d\log\chi^2/ds$ term in Eq.~(\ref{eq:theta:inner}) for small $s$ gives, to zeroth order in $\ve$, 
\begin{align}
s^2 \f{d^2\theta}{ds^2} + s \f{d\theta}{ds} - \theta = 0 {,}
\end{align}
which has a solution $\theta=\alpha s$ satisfying the inner boundary condition.  This confirms the self-consistency of the assumption.    

\subsection{Analytical expression for the Berry curvature} 

The uniform approximation to $\chi(q)$ derived in the previous section provides a compact and physically intuitive expression that accurately incorporates nonadiabatic effects near the conical intersection.  It has proved difficult to derive a similar approximation for $\theta(s)$ due to the nonlinearity of its differential equation.  Moreover, since $\theta(s)$ is defined in terms of the ratio of two small quantities, $a(s)$ and $b(s)$, it is also challenging to approximate starting from the linear equations.  In this section, we propose a one-parameter approximation that provides an accurate fit to $\theta(s)$ over a range of $\epsilon$.  

The approximation we propose is 
\begin{align}
1-\cos\theta(s) = (1+(s/s_0)^{-\nu})^{-\mu} {.} \label{eq:oneminuscos}
\end{align} 
To determine the parameters, we require that $\theta(s)$ has the correct local behavior $\theta(s)\sim \alpha s$ as $s\rightarrow 0$, where $\alpha$ was related to the energy eigenvalue in the previous section.  This implies $\mu\nu=2$ and $s_0=\sqrt{2}/\alpha$.  Setting $\mu=2/\nu$, the one remaining parameter $\nu$ has been determined as a function of $\ve$ by fitting Eq.~(\ref{eq:oneminuscos}) to the numerically exact solution.  The resulting $\nu(\epsilon)$ is a slowly varying function of $\ve$ that can be accurately fit by 
\begin{align}
\nu = \f{2.436 + 0.225\, \ve^{1/2}}{1+ 0.124\, \ve^{1/2}}
\end{align}
over the range $\ve = (0.01,0.5)$.
The above approximation for $\theta(s)$ directly determines the Berry phase in Eq.~(\ref{eq:gamma:JT}).  The corresponding Berry curvature in Eq.~(\ref{eq:Berrycurvature}) is
\begin{align*}
B_{Q_2Q_3} = \f{\hbar}{Q_0^2 \ve^2} \f{1}{ss_0} \big( 1+ (s/s_0)^{-\nu} \big)^{-1-\mu} (s/s_0)^{-1-\nu} {.}
\end{align*} 
As expected, the Berry curvature is localized at the origin and its width in terms of the variable $s$ is $s_0=\mathcal{O}(1)$.  Translated back to the original coordinate $Q$, this implies a width of order $\hbar \K^{\f{1}{2}}/g \M^{\f{1}{2}}$, as anticipated in Eq.~(\ref{eq:Qsmear}) and confirming the numerical analysis of Ref.~\onlinecite{requist2016a}.

\section{\label{sec:surface} Exact potential energy surface}

An important product of the exact electron-nuclear factorization \cite{hunter1975,gidopoulos2014,abedi2010} is the derivation of a potential energy surface $\mathcal{E}(R)$ which is exact in the sense that when it is used together with the induced vector potential $A_{\mu}(R)$ in the nuclear Schr\"odinger equation, the solution reproduces the nuclear wave function of the exact electron-nuclear factorization.  
Here, we examine the nonadiabatic contributions to this exact potential energy surface in the large mass limit of the linear $E\otimes e$ Jahn-Teller model.  

\begin{figure}[t]
\centering
     \includegraphics[width=\columnwidth]{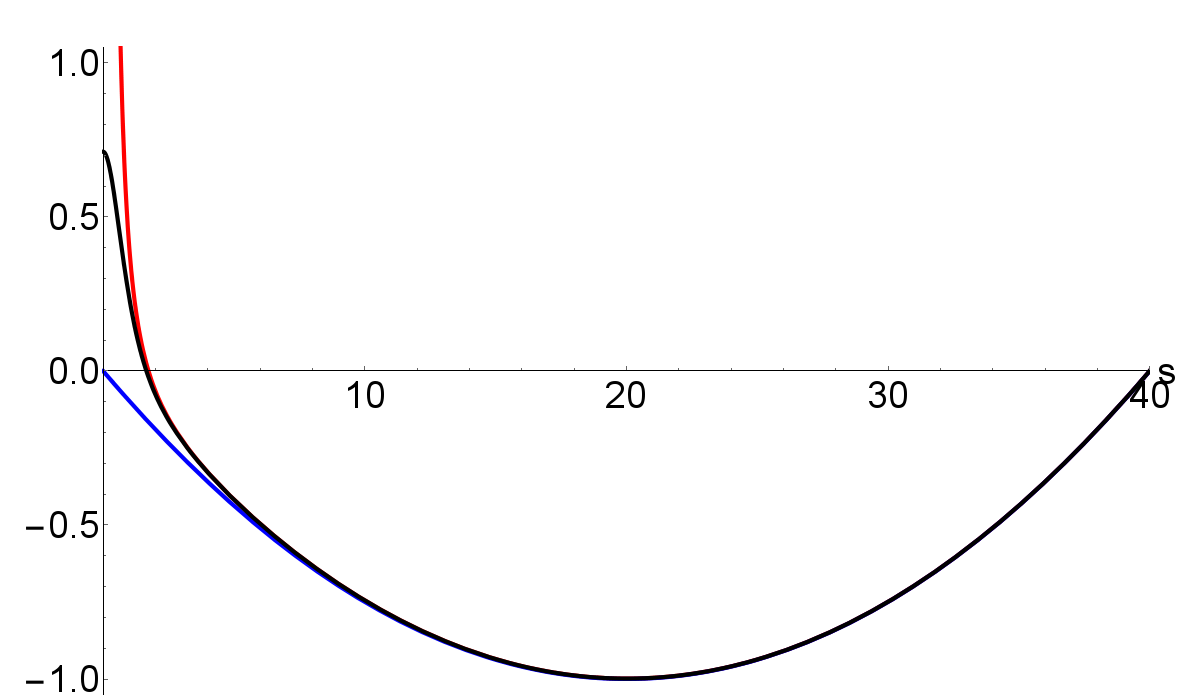} 
     \caption{The effective potential energy surface $\mathcal{E}_{\rm eff,in}(s)$ [black] and the BO potential energy surface with and without the centrifugal potential [red and blue] are plotted for $\epsilon=1/20$.}
     \label{fig:EpsilonEff}
\end{figure}

Nonadiabatic effects enter solely through the $\theta$ dependence of the exact potential energy surface.  Since these effects are localized near the origin, we will focus on the effective one-dimensional potential $\mathcal{E}_{\rm eff,in}(s)$ that appears in the differential equation for $\chi(s)$ in the inner region, Eq.~(\ref{eq:chi:inner}).  $\mathcal{E}_{\rm eff,in}(s)$ is the sum of $\mathcal{E}(s)$ and the centrifugal repulsion $\sin^4(\theta/2)/s^2$, which originates from the vector potential.  If we had an uncoupled nuclear equation with angular momentum quantum number $l$, the centrifugal repulsion would be $l^2/s^2$.  However, as mentioned above, in the $E\otimes e$ Jahn-Teller model with $l=1/2$ the coupling causes a transfer of angular momentum from nuclei to electrons as $s\rightarrow 0$.  The resulting $s$-dependent nuclear angular momentum $L_z(s)/\hbar=\sin^2(\theta/2)$ quenches the divergent centrifugal potential as $s\rightarrow 0$, as seen in the exact surface (black curve) in Fig.~\ref{fig:EpsilonEff}.  In the BO approximation, $\theta=\pi/2$ and there is no quenching (red curve).

A second nonadiabatic effect is the smoothing of the nonanalytic cusp associated with the conical intersection.  The BO potential energy surface without the centrifugal repulsion, the blue curve in Fig.~\ref{fig:EpsilonEff}, shows the characteristic linear dependence near the origin.  In the exact surface $\mathcal{E}_{\rm eff,in}(s)$, the $\sin\theta$ factor multiplying the bare electronic-vibrational coupling $2\ve s$ changes the linear behavior to a regular quadratic behavior, since $\theta\sim\alpha s$.

\begin{figure}[t]
\centering
 \includegraphics[width=\columnwidth]{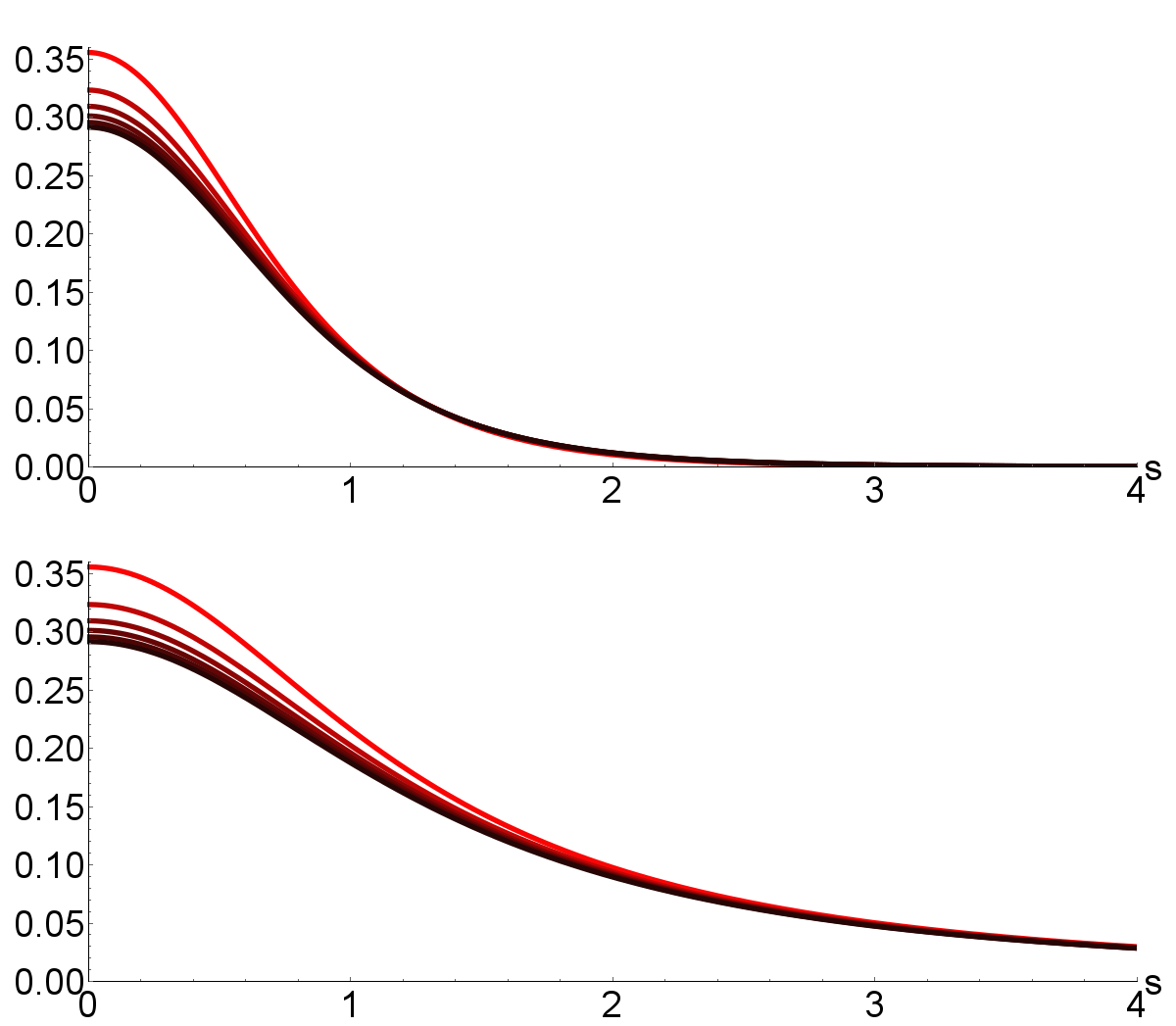}
     \caption{The geometric contributions $\mathcal{E}_{\rm geo,1}$ (top panel) and $\mathcal{E}_{\rm geo,2}$ (bottom panel) to the effective potential energy surface are plotted for the same series of $\ve$ values as in Fig.~\ref{fig:dlogchisq}.}
     \label{fig:EpsilonGeo}
\end{figure}

The remaining nonadiabatic contribution is the following term of geometric origin \cite{requist2016a}, which is responsible for the additional peak in the exact surface near $s=0$:
\begin{align}
\mathcal{E}_{\rm geo}(s) = \f{1}{4} \LB \f{d\theta}{ds} \RB^2 + \f{\sin^2\f{\theta}{2}}{s^2} {,}
\end{align}
We will denote the first term as $\mathcal{E}_{\rm geo,1}(s)$ and the second term as $\mathcal{E}_{\rm geo,2}(s)$.  These contributions are plotted for a series of $\ve$ values in Fig.~\ref{fig:EpsilonGeo}.  According to the definition in Eq.~(\ref{eq:EpsilonGeo}), the geometric term $\mathcal{E}_{\rm geo}(Q)$ vanishes as $\M^{-1}$ as $\M\rightarrow\infty$.  Instead, $\mathcal{E}_{\rm geo,1}(s)$ and $\mathcal{E}_{\rm geo,2}(s)$ are seen to approach universal functions since $\mathcal{E}_{\rm geo}(s) = \ve^{-2} \mathcal{E}_{\rm geo}(Q)$.  In fact, $\mathcal{E}_{\rm geo,1}(s)$ and $\mathcal{E}_{\rm geo,2}(s)$ have the same $s=0$ intercept equal to $\alpha^2/4$, though $\mathcal{E}_{\rm geo,2}(s)$ decays more slowly.  

\section{\label{sec:Ham} Ham reduction factors}

When the induced vector potential $A_{\mu}(R)$ is calculated with the exact conditional electronic wave function, the path-dependent geometric phase $\gamma=\oint A_{\mu} dR_{\mu}$ is an exact and proper gauge-invariant quantity, but it remains an open question to identify experiments that can differentiate it from $\gamma^{\rm BO}$.
Here, we show that Ham reduction factors \cite{ham1965,ham1968,mcconnell1961}, which have long been used to explain the vibronic coupling-induced weakening of the response of Jahn-Teller systems to external perturbations such as magnetic fields, spin-orbit coupling and strain, can be expressed as integrals of the exact geometric phase weighted by the nuclear probability density.  This provides a way to infer the difference between $\gamma$ and $\gamma^{\rm BO}$. 

To see how electronic-vibrational coupling weakens the response of the $E\otimes e$ Jahn-Teller model to external perturbations, first consider the uncoupled problem.  By assumption, the uncoupled electronic states are assumed to be degenerate and transform as an irreducible representation $E$ of the symmetry group, which may be e.g.~the $D_{3h}$ group of a triatomic molecule or the octahedral group $O_h$ of a bulk transition metal impurity.

In the absence of electronic-vibrational coupling, the action of a general perturbation on the electronic states of $E$ symmetry can be represented as
\begin{align}
\hat{V} = V_0 \hat{I} + \vec{V} \cdot \hat{\vec{\sigma}} 
\end{align}
in the basis $\{ |u\rangle, |g\rangle \}$ of electronic states; $\hat{\vec{\sigma}}$ are the Pauli matrices.  The physical effect of the perturbation is fully described by the matrix elements  
$\langle \alpha | \hat{V} | \beta \rangle$; $\alpha,\beta=u,g$. 

When the electronic-vibrational coupling is turned on, the electronic states $|u\rangle$ and $|g\rangle$ evolve into vibronic states $|\Psi_u\rangle$ and $|\Psi_g\rangle$ with the same symmetry as the original electronic states.  Therefore, the coupling preserves the symmetry, but now the matrix elements describing the response of the system to the external perturbation $\hat{V}$ need to be calculated with respect to $|\Psi_u\rangle$ and $|\Psi_g\rangle$.  Since the vibronic wave functions contain electronic and vibrational parts, these matrix elements are reduced in magnitude with respect to the corresponding purely electronic matrix elements.  The action of the perturbation on the vibronic states of $E$ symmetry is therefore
\begin{align}
\hat{V} = V_0 \hat{I} + q V_1 \sigma_1 + p V_2 \hat{\sigma}_2 +  q V_3 \sigma_3 {,} 
\end{align}
where the reduction factors are defined by
\begin{align}
p = \f{\langle \Psi_u | \hat{\sigma}_2 | \Psi_g \rangle}{\langle u | \hat{\sigma}_2 | g \rangle}
\end{align}
and 
\begin{align}
q = \f{\langle \Psi_u | \hat{\sigma}_1 | \Psi_g \rangle}{\langle u | \hat{\sigma}_1 | g \rangle} = \f{\langle \Psi_u | \hat{\sigma}_3 | \Psi_u \rangle}{\langle u | \hat{\sigma}_3 | u \rangle} {.}
\end{align}
In the linear $E\otimes e$ Jahn-Teller model, $p$ and $q$ are related by the identity $q = (1+p)/2$ \cite{ham1968}.

It is now simple to show that in the linear $E\otimes e$ Jahn-Teller model $p$ and $q$ can be expressed in terms of the exact geometric phase in Eq.~(\ref{eq:gamma:JT}).  In the notations of Sec.~\ref{sec:Jahn-Teller}, the expressions for $p$ and $q$ become
\begin{align}
p &= 2\pi \int_0^{\infty} dQ Q |\chi(Q)|^2 \cos\theta(Q) \nn \\
q &= 2\pi \int_0^{\infty} dQ Q |\chi(Q)|^2 \f{1+\cos\theta(Q)}{2} {.}
\end{align}
Since $\cos\theta(Q)$ is proportional to the conditional electronic angular momentum $l_z(Q)$, $p$ will be small if the electronic angular momentum is effectively quenched at all values of $Q$ for which $|\chi(Q)|^2$ is appreciable.  This is clearly the case for the states shown in Fig.~\ref{fig:chitheta-vs-Q}, as $\cos\theta(Q)$ is only significantly different from zero in the classically forbidden region near the origin where $|\chi(Q)|^2$ is exponentially small.  Using the expression for the exact geometric phase in Eq.~(\ref{eq:gamma:JT}), $p$ can be expressed as
\begin{align}
p = \int_0^{\infty} dQ Q |\chi(Q)|^2 \Big( 1 - \f{\gamma(Q)}{\pi} \Big) {.}
\end{align}
The more rapidly $\gamma(Q)$ saturates to its asymptotic value, i.e.~the more localized the Berry curvature, the smaller the value of $p$.  In the BO limit, $p=0$.

\section{Conclusions}

The adiabatic molecular Berry phase depends on the nonanalyticity implied by conical intersections of BO potential energy surfaces.  Yet points of conical intersection are precisely where the BO approximation breaks down most severely, raising doubts about whether the molecular Berry phase would survive in an exact calculation.  In fact, an example was found in which the molecular Berry phase becomes identically zero when calculated with the conditional electronic wave function from the exact factorization scheme instead of the BO wave function \cite{min2014}.  Hence, the adiabatic molecular Berry phase is in this case an artifact of the BO approximation.  

Spectroscopic signatures of the Berry phase have been observed in Jahn-Teller systems since the 1960's \cite{ham1987,ham1972,englman1972}.  Although the BO approximation breaks down at conical intersections in these systems, the effects of the Berry phase are nevertheless observable because they influence the global behavior of the conditional electronic wave function far from the point of conical intersection.  However, the specific {\it topological} character of the adiabatic Berry phase in Jahn-Teller systems is not a true and observable feature of the exact wave function \cite{requist2016a}.

When the Berry phase is calculated with the exact conditional electronic wave function in Jahn-Teller systems, it becomes a genuinely path-dependent quantity that is close to but slightly less than $\pi$ for most paths.  The deviation from $\pi$ is a nonadiabatic effect that arises because the Berry curvature---a featureless Dirac delta function in the BO approximation---gets broadened into a smooth peaked function in an exact calculation based on Eq.~(\ref{eq:geomphase:exact}).  
That the Berry phase is close to $\pi$ follows from the fact that the Berry curvature is highly localized so that all but the smallest paths pick up most of the weight of the peak and thus almost recover the adiabatic result.  In physical terms, the breakdown of the BO approximation at conical intersections has only a small perturbative effect on the exact conditional electronic wave function at faraway points, and since the wave function at those points is therefore close to the BO wave function, the value of the Berry phase calculated on a path that stays away from the conical intersection is close to its BO value.

The precise value of the Berry phase for a given path depends on the detailed shape and extent of the Berry curvature.  One of the main objectives of this paper was to derive an analytical formula that accurately describes the Berry curvature in the large mass limit of the prototypical linear $E\otimes e$ Jahn-Teller model.  Although we have found numerically that the Berry curvature approaches a universal function in the limit $\M\rightarrow 0$, we were not able to find its analytical form in terms of special functions.  Nevertheless, we have proposed a compact formula that we hope will prove helpful in designing functional approximations in a nonadiabatic generalization of density functional theory, in which the exchange-correlation energy depends on the Berry curvature \cite{requist2016b}. 

Since we cannot force the nuclei to move along any given path, the exact molecular Berry phase can only be inferred from an observable that will involve an integral over nuclear configuration space of a parametrically $R$-dependent conditional variable weighted by the nuclear probability density.  It has been an open question to identify an experimental observable that clearly distinguishes between the molecular geometric phase calculated with the exact conditional electronic wave function from that calculated with the BO wave function.  We have shown here that Ham reduction factors, which describe e.g.~how electronic-vibrational coupling modifies $g$-factors in electron spin resonance experiments, are related to an integral over the exact molecular geometric phase.  Thus, experimental measurements of Ham reduction factors are sensitive to the difference between the exact and adiabatic molecular Berry phases.

{\it Note added.} Two articles relevant to dynamical Jahn-Teller effects and Berry phases have recently appeared.  Ribeiro and Yuen-Zhou explain the reason for ground state degeneracies in Jahn-Teller models with maximal continuous symmetries \cite{ribeiro2017}.  Thiering and Gali present {\it ab initio} calculations for the dynamical Jahn-Teller induced damping (Ham effect) of the spin-orbit interaction in diamond nitrogen-vacancy centers \cite{thiering2017}.

\bibliography{bibliography}

\end{document}